\documentclass[preprint,nofootinbib,longbibliography,groupedaddress]{revtex4-2}
\usepackage{natbib}
\usepackage{amsmath} 
\usepackage{amssymb}
\usepackage{graphicx}
\usepackage{color} 
\usepackage{url}
\usepackage{tcolorbox}
\usepackage{hyperref}
\tcbuselibrary{fitting,raster,skins,xparse}

\begin{document}
\title{Non-perturbative instanton effects in the quartic and the sextic double-well potential by the numerical bootstrap approach}
\author{Wei~Fan}
\author{Huipeng~Zhang}
\affiliation{Department of Physics, School of Science, Jiangsu University of Science and
Technology, Zhenjiang 212114, China}
\date{\today}
\begin{abstract}
    Recently the non-perturbative numerical bootstrap method is  rapidly developing, especially its application in nonlinear quantum mechanics, many-body physics, lattice models and matrix models. Here we use this numerical bootstrap to study the non-perturbative instanton effects in symmetric double-well potentials, using an efficient implementation of the algorithm proposed recently by Aikawa, Morita and Yoshimura. The ground state level splitting, caused by instantons, is computed for  the quartic and the sextic case, where the coupling constant $g$ characterizes the strength of  instanton effects. Inspired by well-known perturbative results that is only valid at large $g$ value  under the dilute-gas limit, a qualitative formula is proposed for the ground state level splitting across all values of $g$, which agrees well with the bootstrap data. It has the expected behavior at both large value $|g|\to \infty$ and small value $|g|\to 0$, so it describes both the 'weak' and the 'strong' regimes of instanton effects. This qualitative formula is non-perturbative and can not be obtained by loop summation alone. It is beyond the dilute-gas approximation, and might be from a renormalization-group-like procedure or other undiscovered analytic methods. 
    
\end{abstract}

\maketitle
\tableofcontents

\section{\label{sec:intro}Introduction}

The instanton physics plays an important role in gauge theory~\cite{Belavin:1975fg, tHooft:1976rip,Jackiw:1976pf,Callan:1976je}, quantum tunneling~\cite{Polyakov:1976fu}, vacuum decay~\cite{Callan:1977pt} and bubble nucleation~\cite{Coleman:1980aw}. Its effects are perturbatively computed via loop expansions using path integrals.  The double-well potential~\cite{PhysRev.184.1231,*PhysRevLett.27.461,Gildener:1977sm} serves as the prototypical instanton problem~\cite{Polyakov:1976fu}. For the quartic double-well, the instanton physics permits a semiclassical interpretation of the tunneling process that leads to the level splitting of the two degenerate energy eigenstates. Loop expansions (or WKB methods) show that the level splitting decreases exponentially~\cite{Coleman:1985rnk,Muller-Kirsten:2012wla,Kleinert:788154} with the increasing barrier height (the difference between the local maximum and the global minimum of the double-well potential and is proportional to  the coupling  constant $g$ in our notation). This effect has been computed up to the two-loop level~\cite{Kleinert:788154} under the dilute-gas limit.

An interesting question to ask is what \emph{the non-perturbative instanton effects}~\footnote{Here perturbation refers to loop expansions in path integral, not the large order perturbation expansions of ordinary quantum mechanics.} would be. This paper will address this question using the numerical bootstrap method.  The bootstrap is an important topic in quantum field theory, where the analytic bootstrap usually involves very complicated mathematical techniques~\cite{Kruczenski:2022lot,Hartman:2022zik,Bissi:2022mrs}. In quantum mechanical bootstrap, the theory's spectrum can be solved from a set of consistency conditions, which are constructed only using elementary properties of quantum theory, so no perturbation is referred to in this approach.  The  numerical bootstrap gets a great improvement after incorporating the semidefinite programming  algorithm~\cite{Poland:2011ey,Kos:2013tga,Kos:2014bka,Simmons-Duffin:2015qma}, and this leads to many important results~\cite{Poland:2018epd,Poland:2022qrs}, for example, the high precision value of critical points of the 3D Ising model. It also leads to a great development of the numerical S-matrix bootstrap~\cite{Paulos:2016fap,*Paulos:2016but,*Paulos:2017fhb}.

Recently the semidefinite programming is adapted to matrix quantum mechanics~\cite{Anderson:2016rcw,Lin:2020mme,Han:2020bkb}, which expands the scope of application of numerical bootstrap. It is now widely~\footnote{This list of literature is still rapidly growing, so here it is by no means a complete one.} used in  nonlinear quantum mechanics~\cite{Berenstein:2021dyf,Berenstein:2021loy,Bhattacharya:2021btd,Aikawa:2021eai,Li:2022prn,Hu:2022keu,Aikawa:2021qbl,Tchoumakov:2021mnh,Bai:2022yfv,Khan:2022uyz,Berenstein:2022ygg,Morita:2022zuy,Blacker:2022szo,Nakayama:2022ahr,Berenstein:2022unr,Guo:2023gfi,Berenstein:2023ppj}, 
quantum many-body physics models~\cite{2020arXiv200606002H,Lawrence:2021msm,Nancarrow:2022wdr} and high energy physics models~\cite{Hessam:2021byc,Kazakov:2021lel,Du_2022,Lin:2023owt}. 

In this paper, we will use the numerical bootstrap to study the non-perturbative instanton effects. The model used here is the symmetric double-well potentials, which usually serves as a prototype of field theory instanton problems. Besides instanton physics, its energy spectrum is itself an important mathematical physics problem~\cite{Caswell:1979qh,Chhajlany_1990, Liang:1992ms,Liang:1992mw,Yukalov_1996,garg2000tunnel,Serone:2016qog, Kreshchuk:2018qpf,Fischbach:2018yiu} involving WKB methods, Picard-Fuchs methods and so on.

For the quartic double-well, the instanton physics is computed for large  $g$ values in numerical bootstrap~\cite{Berenstein:2021loy,Bhattacharya:2021btd}. Their results show that the two-loop perturbation is invalid when $|g|$ becomes small. Only when the coupling  constant $|g|$ is very large (that means large barrier height), the two-loop perturbation agrees well with the non-perturbative bootstrap. This is understood as follows: with a large barrier height the instanton effect is severely prohibitated, so the dilute-gas approximation~\footnote{In the dilute-gas limit, the classical  configuration is composed of instantons that are far away separated from each other with no interactions, which can be viewed as a dilute 'gas' of instantons.} used in perturbative loop expansions becomes accurate and ensures the perturbation. So the numerical bootstrap shows that  perturbative loop expansions fail for the small barrier height, where the instanton effects is active and the dilute-gas approximation is no longer valid. 

We will further analyze the instanton effects in the quartic double-well. We can use its  perturbative two-loop formula~\cite{Kleinert:788154} as a prototype and  propose a qualitative non-perturbative formula~\eqref{eq:formulaH4} for the dependence of ground state level splitting on all values of $g$. It agrees well with bootstrap results and describes  both the 'weak' and the 'strong' instanton regime. Then we will further study the instanton physics in the sextic double-well and justify the qualitative formula~\eqref{eq:formulaH4}. This suggests that for double-well of all anharmonicities the dependence of ground state level splitting on $g$ might have the same functional form~\eqref{eq:formulaH4}, only with different parameter values.  These models have rich dynamics of oscillatons and oscillons~\cite{Gani:2014gxa,Fodor:2019ftc,MoradiMarjaneh:2022vov,Blinov:2022twe} and play an important role in the structual phase transition~\cite{Krumhansl:1975zz,PhysRevB.18.6139,PhysRevLett.72.2130,Khare:2014kva} of ferroelectric materials. 

The paper is organized as follows. In Section~\ref{sec:bootstrap}, we will review steps of the numerical bootstrap method. We will use a  more efficient implementation of the numerical bootstrap proposed in~\cite{Aikawa:2021qbl}. In Section~\ref{sec:results}, we will compute and discuss the ground-state level splitting obtained from numerical bootstrap for symmetric double-wells.  We will firstly redo the computation for the quartic double-well  using  the efficient implementation~\cite{Aikawa:2021qbl}. Then we propose a qualitative formula for the dependence of ground state level splitting on coupling $g$. After that, we will study  the level splitting behavior of the  sextic double-well and check the validity of the qualitative formula. We close in Section~\ref{sec:conclusion} with a discussion of open questions and future work.

\section{\label{sec:bootstrap}Method of the numerical bootstrap}

In this section we briefly review the method of numerical bootstrap. For the practical implementation of the algorithm, we will follow the choice of operators $\{x^mp^n\}$ introduced firstly by~\cite{Aikawa:2021qbl}. The algorithmic details have been analyzed extensively from various aspects in the literature of numerical bootstrap. So here we only sketch the steps and will not go into the algorithmic details. The interested reader can go to references listed in the Introduction~\ref{sec:intro}. 

The numerical bootstrap divides into three  steps: (1) choose a set of operators and derive their recursive equations, (2)  impose positive constraints (or other working constraints) on the operators and obtain the bootstrap matrix, (3) set the numerical search space and use the semidefinite optimization to find allowed parameter values in this space. 
For double-well potentials, we choose the system energy as the search parameter, because our aim is to compute the ground state level splitting.

\subsection{Recursive equations}

   For a quantum mechanical system, we use the following Hamiltonian:
\begin{equation}\label{one}
        H=\frac{p^2}{2}+V(x)
\end{equation}
For its energy eigenstate  $\left| \varphi  \right\rangle$ with eigenvalue $E$, the expectation value of operators composed of $H$ and an arbitrary operator $\alpha $, must satisfy the following two constrains, 
\begin{equation}\label{two}
       \left\langle {\left[ {H,\alpha } \right]} \right\rangle \equiv \left\langle  \varphi | {\left[ {H,\alpha } \right]} | \varphi \right\rangle =   0, \quad
       \left\langle {H\alpha } \right\rangle  = \left\langle {\alpha H} \right\rangle  = E\left\langle \alpha  \right\rangle 
\end{equation} 
If choose $\alpha  = {x^n}, {x^n}p$, these constrains lead to the following recursion relations~\cite{Aikawa:2021qbl,Hu:2022keu}
\begin{equation}\label{sev}
        n(n - 1)(n - 2)\left\langle {{x^{n - 3}}} \right\rangle  - 8n\left\langle {{x^{n - 1}}V(x)} \right\rangle  + 8nE\left\langle {{x^{n - 1}}} \right\rangle  - 4\left\langle {{x^n}{V'}(x)} \right\rangle  = 0.
\end{equation}

\subsection{bootstrap matrix}
To build the bootstrap matrix, we follow the efficient construction proposed by~\cite{Aikawa:2021qbl} by choosing the following operators,
\begin{equation}\label{ten}
       \mathcal{O}_{x}  = \sum\limits_{n = 0}^{k} {{C_n}{x^n}},        
 \mathcal{O}_{xp}  = \sum\limits_{m = 0}^{k_x}\sum\limits_{n = 0}^{k_p} {{C_{mn}}{x^m}{p^n}}
\end{equation}
with ${C_n}, {C_{mn}}$ being constants. The expectation value of the operator $\mathcal{O}=\mathcal{O}_x, \mathcal{O}_{xp}$ on the energy eigenstate $\left| \varphi  \right\rangle $  satisfies the positivity constraint
\begin{equation}\label{ele}
         \left\langle {{{\mathcal{O}} ^\dag }{\mathcal{O}} } \right\rangle \geq 0.
\end{equation}
From this constraint, the bootstrap matrix can be built that must be positive-semidefinite
\begin{equation}\label{eq:shisan}
        {M}: = \left( {\begin{array}{*{20}{c}}
{{\mathcal{O}}_0^{\dag }{{\mathcal{O}}_0}} & {{\mathcal{O}} _0^{\dag }{{\mathcal{O}} _1}}& \ldots &{{\mathcal{O}} _0^{\dag }{{\mathcal{O}} _k}}\\
{{\mathcal{O}} _1^{\dag }{{\mathcal{O}} _0}}&{{\mathcal{O}} _1^{\dag }{{\mathcal{O}} _1}}& \ldots &{{\mathcal{O}} _1^{\dag }{{\mathcal{O}} _k}}\\
 \vdots & \vdots & \ddots & \vdots \\
{{\mathcal{O}} _k^{\dag }{{\mathcal{O}} _0}}&{{\mathcal{O}} _k^{\dag }{{\mathcal{O}} _1}}& \ldots &{{\mathcal{O}} _k^{\dag }{{\mathcal{O}} _k}}
\end{array}} \right)  \geq 0,
\end{equation}
where $\mathcal{O}_{0,1,2,\ldots}$ are the component operators of $\mathcal{O}$ and the maximum value $k$ is called the depth of the bootstrap matrix. As the depth $k$ increases, the positive-semidefinite constraint becomes stronger and so the numerical results becomes more accurate.

\subsection{Search space}

The last step is to set the search space  which is a minimum set of data to initialize the recursion. After choosing search parameters and the optimization target, the positive-semidefinite optimization will exclude  parameter values that do not satisfy the constraints~\eqref{eq:shisan} and so reduce the size of the parameter space. With sufficiently large depth $k$, the remaining parameter space passing the constraints~\eqref{eq:shisan} will be a tiny neighborhood that can be viewed as a data point. This data point is the discrete eigenvalue of the quantum system with the optimized target value being the expectation. For double-wells, we choose the system energy as the search parameter. The search space of the quartic double-well is $\left\{ E,{\rm{ }}\left\langle {{x^2}} \right\rangle  \right\}$ and of the sextic case is $\left\{ {E,{\rm{ }}\left\langle {{x^2}} \right\rangle ,{\rm{ }}\left\langle {{x^4}} \right\rangle } \right\}$.

\section{\label{sec:results}Bootstraping double-well potentials}

Here we apply the above bootstrap method to compute the ground state level splitting of the quartic and the sextic double-well. We will focus on the physics in the bootstrap results and propose a qualitative formula for the dependence of the level splitting on the coupling $g$. For  algorithmic details, the interested reader can go to the references in Section~\ref{sec:intro} and~\ref{sec:bootstrap}.  

The symmetric double-wells are defined as 
\begin{equation}
\label{eq:Hn}
H=\frac{p^2}{2} + g x^2 + x^{2n}, \quad g < 0 \,\& \, n\geq 2.
\end{equation}
In this convention,  the potential has the following global minimum and local maxmimum 
\begin{equation}
    V\left(x=\pm (\frac{|g|}{n})^{1/(2n-2)}\right) =  - \frac{n-1}{n} (\frac{1}{n})^{2/(2n-2)} |g|^{2n/(2n-2)}, \quad V(x=0)=0. 
\end{equation}
The barrier height $V_m$, which is defined as the difference between the local maximum and the global minimum, is  proportional to the coupling constant $g$ 
\begin{equation}
     V_m = \frac{n-1}{n} (\frac{1}{n})^{2/(2n-2)} |g|^{2n/(2n-2)}.
\end{equation}  

Classically the two global minimum are degenerated. In quantum physics,  instantons connect them and there will be ground state level splitting $\Delta E$. With increasing activity of instantons, the level splitting  $\Delta E$ will also increase. Intuitively an increasing barrier height $V_m$ would suppress the tunneling of instantons, so we can  use the value of $g$ to characterize  the activity of instantons and the ground state level splitting will be a function of the coupling $\Delta E=\Delta E(g)$.

\subsection{\label{sec:quartic}Quartic double-well}

For   large  $g$ values, the ground state level splitting of the quartic double-well has been computed in~\cite{Berenstein:2021loy,Bhattacharya:2021btd}, and they find that only in the asymptotic regime of large barrier height the bootstrap agrees with the perturbation result. Here we  redo the computation using the algorithm review in  Section~\ref{sec:bootstrap}, and use its perturbative results as a prototype of a qualitative formula valid across all values of $g$.

The quartic double-well 
\begin{equation}
\label{eq:H4}
H=\frac{p^2}{2} + g x^2 + x^4, \quad g < 0,
\end{equation}
has the global minimum  $-|g|^2/4$ at $x=\pm \sqrt{|g|/2}$ and local maximum $0$ at the center $x=0$. The barrier height is $V_m=|g|^2/4$.  The perturbative one-loop~\cite{Muller-Kirsten:2012wla,Kleinert:788154}  and two-loop~\cite{Kleinert:788154} results of the ground state level splitting are as following
\begin{equation}
    \label{eq:loop}
    \Delta E^{(1)}=\frac{8|g|^{5 / 4}}{\sqrt{\pi}} \exp \left[-\frac{2}{3}|g|^{3 / 2}\right], \quad  \Delta E^{(2)}=\frac{8|g|^{5 / 4}}{\sqrt{\pi}} \exp \left[-\frac{2}{3}|g|^{3 / 2} -\frac{71}{48|g|^{3 / 2}}\right].
\end{equation}
It is controlled or weighted by the exponential that looks like a Boltzmann weight, where a power of $|g|$ plays the role of energy. This is consistent with the instanton physics that large $|g|$ (and so large barrier height $V_m$) suppresses instantons. 

\begin{figure}
\centering
\includegraphics[width=1\linewidth]{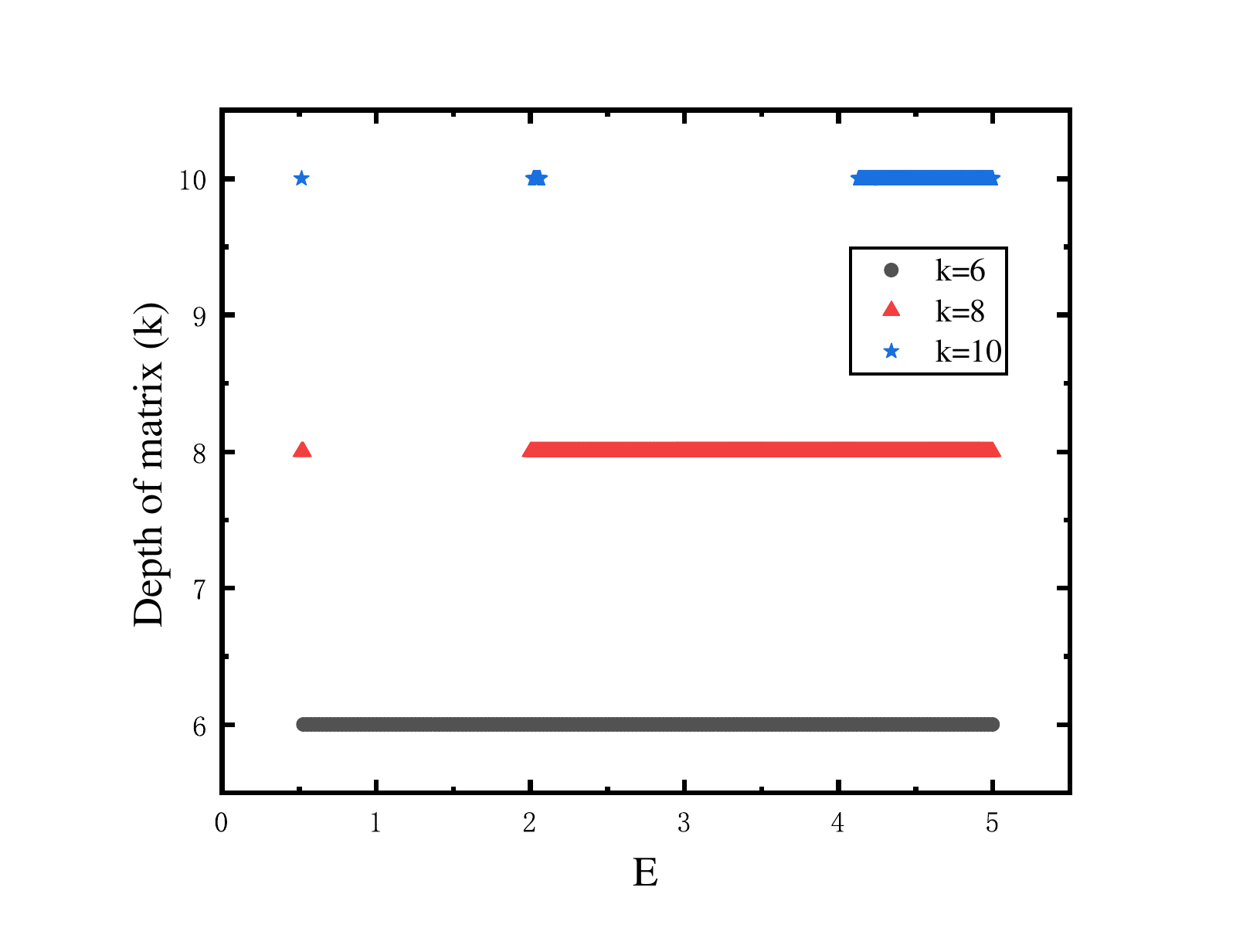}
\caption{An example bootstrap data of the quartic double-well at $g=-1/2$ for various depth $k$ of bootstrap matrix. For small depth $k=6$, the remaining parameter space of energy is a large band. As the depth increases to $k=8$, the ground state energy is identified. When the depth increases further to $k=10$, the first excited energy also get identified.}\label{fig:band4} 
\end{figure}

An example of the bootstrap data is shown in Figure~\ref{fig:band4} for the quartic double-well at $g=-1/2$. It explains the bootstrapping process that when the depth $k$ of bootstrap matrix increases, discrete eigenvalues get identified one by one. The only thing one does is to increase the depth of bootstrap matrix, and this is completely a non-perturbative procedure.  

\begin{figure}
\centering
\includegraphics[width=1\linewidth]{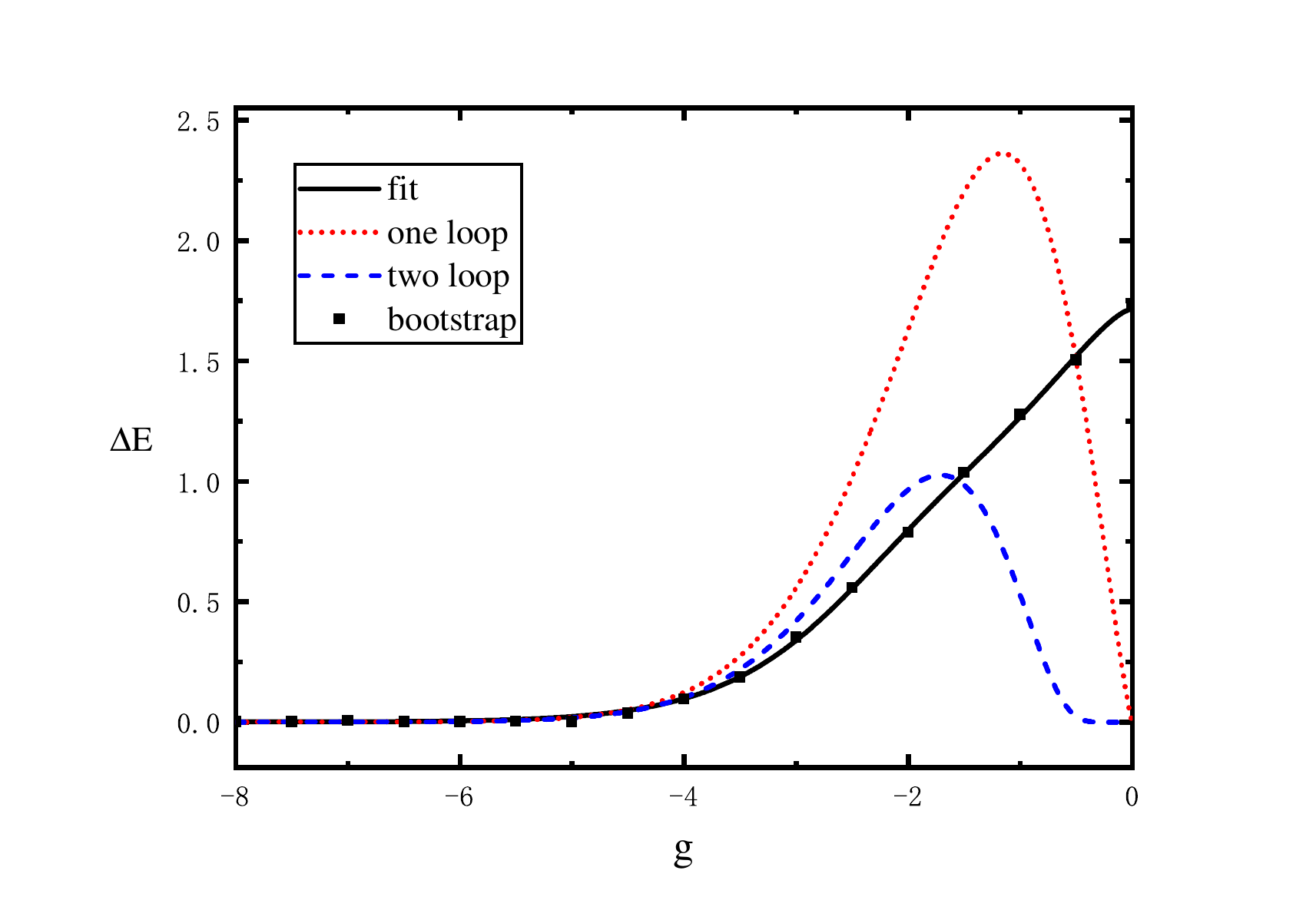}
\caption{The ground state level splitting $\Delta E$ for different $g$ values of the quartic double-well. The squared dots are bootstrap data. The dotted and the dashed line are the perturbative one-loop and two-loop formula. Only for large $|g|$ values where the instantons can be described as a dilute 'gas', the perturbative formulae agree with the bootstrap data, which is consistent with the physics of loop expansions for they are perturbations around the 'weak' regime of instanton effects. For small $|g|$ values where the instanton effects are 'strong', the perturbative formulae do not give the correct description. The solid line 'fit' is the numerically fitted curve for the qualitative formula~\eqref{eq:formulaH4} we proposed to describe both the 'strong' and the 'weak' regime of instanton effects. The numerically fitted parameter values are $A=3.4328685476609424$, $a=0.4195968835224752$, $b=1.5429284622262658$, $c=0.24786203154738526$, $d=2.412365704471155$. The aim of this data fitting is to show the power of the qualitative formula~\eqref{eq:formulaH4} and the true parameter values should be obtained from an analytic method. 
}\label{fig:H4}
\end{figure}

The dependence of $\Delta E$ on $g$ obtained from numerical bootstrap of~\eqref{eq:H4} is shown in Figure~\ref{fig:H4}. The perturbation formulae~\eqref{eq:loop} do not agree with the bootstrap data. Only at very large $|g|$ values the formulae~\eqref{eq:loop} correctly describe $\Delta E$. At large $|g|$, the barrier height $V_m$ is also large and the instantons are suppressed. Then the dilute-gas approximation used in loop perturbation becomes valid. As explained in the Introduction, under this approximation instantons are far away separated from each other with no interactions and can be viewed as a dilute 'gas', so it is a regime of 'weak' instanton effects.
At intermediate $|g|$ values, the loop formulae start to fail,  although the two-loop formula behaves better the one-loop formula. Near $|g|\to 0$, loop expansions give $\Delta E\to 0$ and this is completely wrong. As  $|g|\to 0$, the barrier height is very small and there are very strong activity of instantons, which might be viewed as strongly interacting dense 'gas' (or something else beyond this). Then intuitively, perturbations around configurations without interactions would fail at the regime of 'strong' instanton effects. So the perturbative loop computations are asymptotic expansions around large $|g|$ value, and can only describe 'weak' instanton effects. 

The bootstrap data shows that $\Delta E$ is a well-behaved function of $g$ without any singularities or discontinuities. This suggests that there should be a closed formula of $\Delta E(g)$, if one can find a hypothetical method that is valid at both  the 'weak' and the 'strong' instanton effects to analytically compute $\Delta E$. The perturbation formulae~\eqref{eq:loop} show that the leading contribution is an exponential function of $|g|$, and  the subleading contribution is suppressed by an extra exponential. Based on the above analysis, we propose the following qualitative formula of $\Delta E(g)$ with positive parameter values to be determined by the detailed model
\begin{tcolorbox}[enhanced,colframe=green!20!black,colback=white]
\begin{equation}
    \label{eq:formulaH4}
    \Delta E(g) = A \frac{e^{-a |g|^{b}}}{1+e^{- c|g|^d}}, \quad A, a, b, c, d > 0.
\end{equation}\end{tcolorbox}
The parameter $A= 2 \Delta E(g=0)$ is a normalization determined by the maximum level splitting at $g=0$. The 'weak' instanton effects at large $|g|$ value   suggest  the numerator to be an exponential  that is similar to the Boltzmann weight in the one-loop formula~\eqref{eq:loop}, so it correctly describes the vanishing of $\Delta E$. On the other hand, the 'strong' instanton effects near $|g|\to 0$ suggest the form of the denominator, which is finite at $g=0$ and can be expanded as a well-behaved Taylor series of $|g|$. This is consistent with the Rayleigh–Schrödinger perturbation theory: suppose one can carry out the  perturbation computations for small negative $g$ the result would be a Taylor series of $g$. Actually near $g=0$, the splitting $\Delta E$ is computed to be a polynomial of $g$ up to order $\mathcal{O}(g^3)$ by a variational method~\cite{Yukalov_1996} beyond the Rayleigh–Schrödinger perturbation. So the proposed formula makes sense near $|g|\to 0$. This denominator is \emph{non-perturbative} and can not be obtained from naive loop summation alone, because the loops are expanded around the dilute-gas configurations at $|g|\to \infty$ and they can not probe the strongly interacting instanton physics at $|g| \to 0$. This denominator may come from a renormalization-group-like procedure. 

We fit this qualitative formula~\eqref{eq:formulaH4} of $\Delta E(g)$ to the bootstrap data and show it Figure~\ref{fig:H4}. It agrees well with the data and the numerical values of parameters are listed there. The value $b=1.5429284622262658$ of the Boltzmann weight in the numerator is close to the value  $3/2$ of the one-loop formula~\eqref{eq:loop}. The value $d=2.412365704471155$ of the exponential in the denominator is severely different from the value $-3/2$ in the extra exponential introduced by the two-loop computation~\eqref{eq:loop}, which shows that the denominator is not from a naive loop summation.

Here we emphasize that, the sole purpose of this data fitting is to justify  the qualitative formula~\eqref{eq:formulaH4}. The true  values of parameters $a, b, c, d$ should be determined analytically, maybe from a perturbation combined with renormalization-group. The numerically fitted values listed in the Figure should not be taken as something exact or something serious.

\subsection{\label{sec:pureSextic}Sextic double-well}

Now we do the numerical bootstrap for the sextic double-well and check the validity of qualitative formula~\eqref{eq:formulaH4} of $\Delta E(g)$. The sextic double-well
\begin{equation}
\label{eq:H6}
H=\frac{p^2}{2} + g x^2 + x^6, \quad g < 0,
\end{equation}
 has the global minimum$-2 |g| \sqrt{|g|}/3\sqrt{3}$ at $x=\pm (|g|/3)^{1/4}$ and local maximum $0$ at the center $x=0$. The barrier height is $V_m=2 |g| \sqrt{|g|}/3\sqrt{3}$. Here there is no perturbation formula for the ground state level splitting, so we will directly fit it to the formula~\eqref{eq:formulaH4}.

\begin{figure}
\centering
\includegraphics[width=1\linewidth]{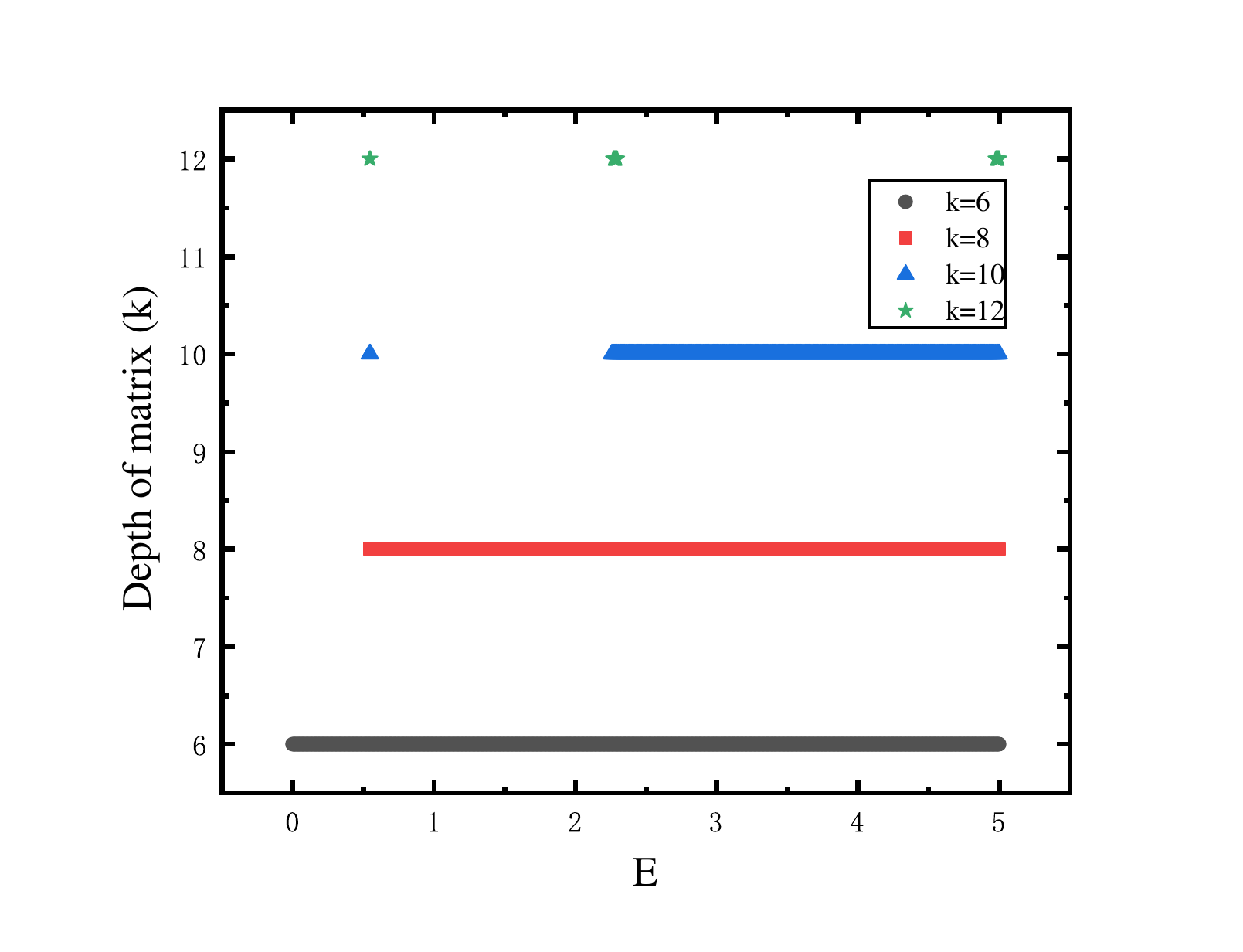}
\caption{An example bootstrap data of the sextic double-well at $g=-1/2$ for various depth $k$ of bootstrap matrix. For small depth $k=6,8$, the remaining parameter space of energy is a large band. As the depth increases to $k=10$, the ground state energy is identified. When the depth increases further to $k=12$, the first excited energy also get identified.}\label{fig:band6} 
\end{figure}

An example of the bootstrap data is shown in Figure~\ref{fig:band6} for the sextic double-well at $g=-1/2$. As the depth $k$ of bootstrap matrix increases, discrete eigenvalues get identified one by one. Here we need the depth $k=12$ to identify the first two eigenvalues, while in the quartic case Figure~\ref{fig:band4} we only need $k=10$. So as the anharmonicity increases, larger bootstrap matrix are required. 

\begin{figure}
\centering
\includegraphics[width=1\linewidth]{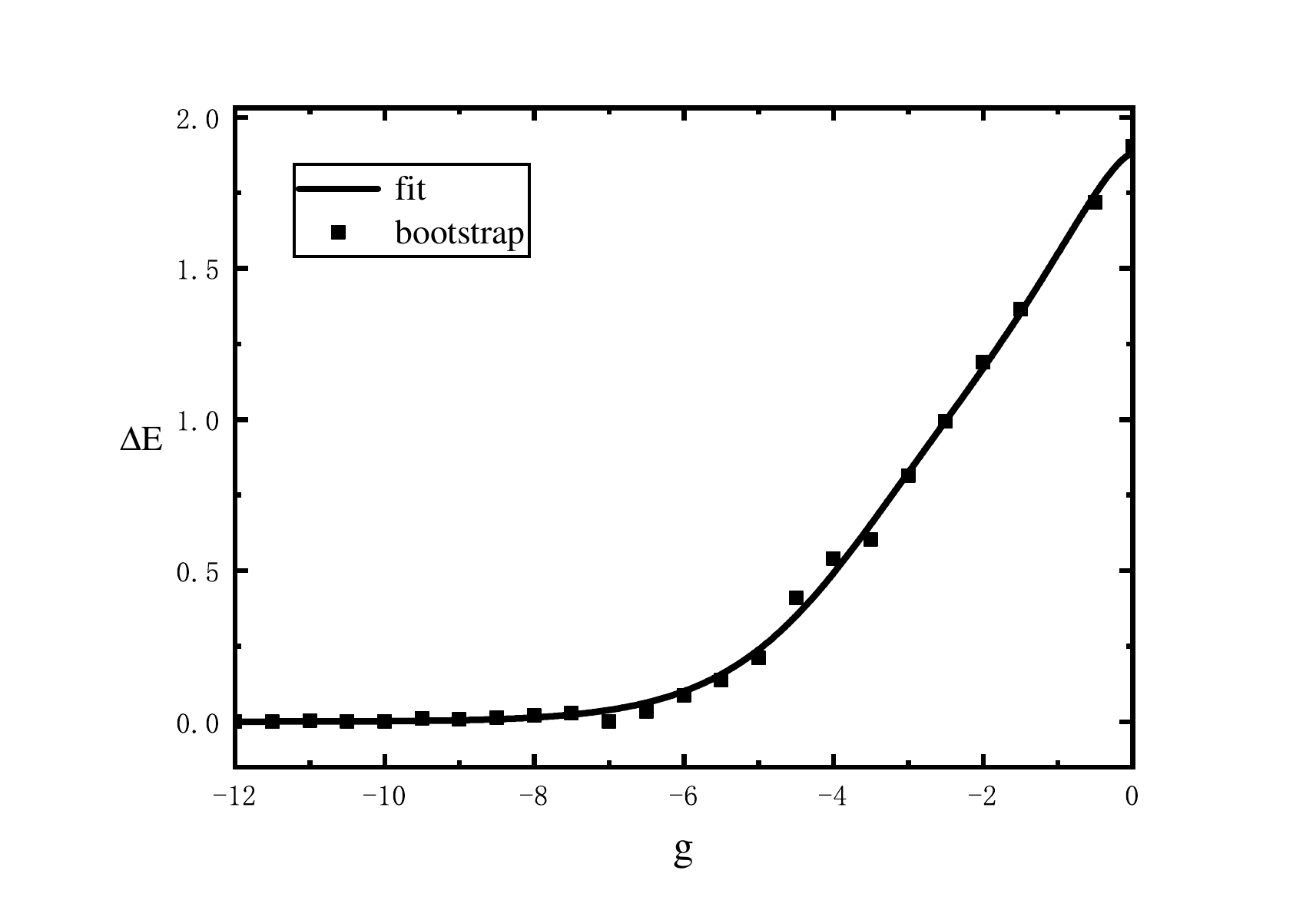}
\caption{The ground state level splitting $\Delta E$ for different $g$ values of the sextic double-well. The squared dots are bootstrap data. There is no loop expansion formula to compare with here.  The solid line 'fit' is the numerically fitted curve for the qualitative formula~\eqref{eq:formulaH4} and it describes  both the 'strong' and the 'weak' regime of the bootstrap data. The numerically fitted parameter values are $A=3.7615670641265995$, $a=0.2360983437455008$, $b=1.5237610476898749$, $c=0.08626481336428367$, $d=2.411061020206574$. Again the aim of data fitting is to justify the  proposed  formula~\eqref{eq:formulaH4} and the true parameter values should be obtained from an analytic method. }\label{fig:H6}
\end{figure}

Figure~\ref{fig:H6} shows the dependence of $\Delta E$ on $g$ obtained from numerical bootstrap of \eqref{eq:H6}. It is similar to the result in Figure~\ref{fig:H4}. Here $\Delta E$ goes to zero at larger values of $|g|$ than the quartic case, because here $V_m \propto  |g|^{3/2}$ is slower than the quartic case $V_m \propto  |g|^{2}$. Here the maximum value of $\Delta E$ is also bigger than the quartic case. For the sextic double-well, there is no loop perturbations of $\Delta E$ to compare with. So we directly fit the qualitative formula~\eqref{eq:formulaH4} with the bootstrap data and show it in Figure~\ref{fig:H6}. It agrees well with the data and the numerical values of parameters are listed there.   Again the focus should be on the validity of the qualitative formula~\eqref{eq:formulaH4}, that describes both the 'weak' and the 'strong' instanton effects.  The numerically fitted values listed in the Figure should not be taken as something exact or something serious.

\section{\label{sec:conclusion}Conclusion}

In this work, we studied  non-perturbative instanton effects in symmetric double-well potentials using the numerical bootstrap method. We followed~\cite{Aikawa:2021qbl} for the detailed implementation of the algorithm. The ground state level splitting $\Delta E$, caused by instantons, was computed for the quartic and the sextic double-well with  values of coupling constant $g$ up to $0$. Inspired by perturbative loop formulae that only work at large coupling $g$, we proposed a qualitative formula~\eqref{eq:formulaH4} for $\Delta E(g)$. This formula works for both large and small $g$ values, so it describes the physics both at 'weak' and 'strong' instanton effects. It is tested on the quartic and the sextic double-well and agrees well with the bootstrap data. The results suggest that the qualitative formula~\eqref{eq:formulaH4} of $\Delta E(g)$ works for symmetric double-well of all anharmoncities~\eqref{eq:Hn}, and different anharmonicities only affect the detailed parameter values. 

The validity of our proposed formula is based on the bootstrap data. One can also obtain the data by numerically integrating the Schrödinger equation, provided that one carefully chooses the integrators and tunes the boundary conditions. Here in this work, we choose the numerical bootstrap method because of its non-perturbative nature. The only artificial numerical ingredient is to increase the depth of the bootstrap matrix. 

Our proposed formula~\eqref{eq:formulaH4} is a non-perturbative formula that can not be obtained from loop expansions. We hope that this formula can serve as a prototype for \emph{finding an analytic method that works both at the 'weak' and the 'strong' regime} of instanton physics. The 'weak' regime is well understood, that instantons are far away separated from each other without interactions and this configuration can be viewed as a dilute-gas. This is well described by the perturbative loop expansions. On the other hand, the 'strong' regime is very different: instantons are very active and are strongly interacting, so they may form a strongly interacting dense 'gas' or even go through a phase transition into a 'solid' phase. This is very obscure and we do not know how to describe this configuration of the 'strong' regime. Obviously traditional loop expansions can not describe this 'strong' regime, because loops are expanded around the dilute-gas configurations which is wrong in the 'strong' regime. 

So new method is needed to compute the instanton effects near the 'strong' regime $|g|\to 0$. Maybe integrability methods or holographic methods are needed, similar to the strong coupling problems of condensed matter physics and gauge theory. By observing the functional form of the proposed formula and the fitted numerical parameter values of the quartic and sextic double-wells, we suggests that the qualitative formula might come from a renormalization-group-like procedure. 

The above discussion of  open problems is for instantons in the path integral approach. The level splitting $\Delta E$ around the 'weak' regime $|g|\to\infty$ can also be computed by WKB methods of quantum mechanics. Within the quantum mechanics approach,  the 'strong' regime $|g|\to 0$ is again a mathematical open problem. Naively by Rayleigh–Schrödinger perturbation theory, $\Delta E$ should be a Taylor serier of $|g|$. But the 'bare' state of $|g|=0$ is unknown,  the naive perturbation can not be carried out in practice. There has been approaches trying variational methods and Picard-Fuchs methods, they finally reduce to differential equation problems, which can only be solved numerically or in asymptotic expansions. Within quantum mechanics,  new method is also needed near the 'strong' regime. 

We hope that the qualitative formula can serve as a guide for searching the analytic method near the 'strong' regime $|g|\to 0$, whether in path integral approach or in quantum mechanics approach. It would be ideal if an analytic method can be discovered that works at both the 'strong' and the 'weak' regime. The double-well potentials also play an important role in kink dynamics and interactions, so our results might also be connected with them.

\begin{acknowledgments}
    Wei Fan is supported in part  by the National Natural Science Foundation of China  (Grant No.12105121). Any opinions, findings, and conclusions or recommendations expressed in this material are those of the authors and do not necessarily reflect the views of the National Science Foundation of China. 
\end{acknowledgments}

\bibliography{boot-instanton}%
\bibliographystyle{apsrev4-2}

\end{document}